\documentclass[aps,
twocolumn,superscriptaddress,amsmath,amssymb]{revtex4}
\usepackage{graphicx}

\begin{document}

\title{Hardy's Paradox and Measurement-disturbance Relations}

\author{Kazuo Fujikawa}
\affiliation{Mathematical Physics Laboratory, RIKEN Nishina Center, Wako 351-0198, Japan}
\author{C.H. Oh}
\affiliation{Centre for Quantum Technologies, National University of Singapore, Singapore 117543, Singapore}
\author{Sixia Yu}
\affiliation{Centre for Quantum Technologies, National University of Singapore, Singapore 117543, Singapore}

\begin{abstract}
We establish a quantitative relation between Hardy's paradox and the breaking of uncertainty principle in the sense of measurement-disturbance relations in the conditional measurement of non-commuting operators. The analysis of the inconsistency of local realism with entanglement by Hardy is simplified if this breaking of measurement-disturbance relations is taken into account, and a much simplified experimental test of local realism is illustrated in the framework of Hardy's thought experiment. 
 The essence of Hardy's model is identified as a combination of two conditional measurements, which give rise to definite eigenvalues to two non-commuting operators simultaneously in hidden-variables models. Better understanding of the intimate interplay of entanglement and measurement-disturbance is crucial in the current discussions of Hardy's paradox using the idea of weak measurement, which is based on a general analysis of measurement-disturbance relations.

\end{abstract}
\maketitle

\section{Introduction}
A thought experiment~\cite{hardy1} and an explicit theoretical 
model~\cite{hardy2} of Hardy propounded a paradox which characterizes Einstein-Podolsky-Rosen phenomena without referring to Bell and Clauser-Horne-Shimony-Holt (CHSH) inequalities~\cite{bell, chsh}. The thought experiment is based on an appealing experimental setting and it motivated many experimental analyses~\cite{togerson} -~\cite{vallone}. In one of the experimental tests of the paradox~\cite{imoto}, the conditional probability such as $P(U_{1}|D_{1}D_{2})$, namely, first measure $D_{1}D_{2}$ and then measure $U_{1}$, appeared as an essential ingredient; here we use the notational conventions of the theoretical model of Hardy~\cite{hardy2} which is explained in detail later in Section 3. The important property to be mentioned here is that $D_{1}$ and $U_{1}$ are not commuting. It is known that the conditional measurement of two non-commuting operators is closely related to uncertainty principle in the sense of measurement-disturbance relations. Also, Hardy's paradox is often discussed in connection with the idea of "weak measurement"~\cite{aharonov1}, and the general analysis of measurement-disturbance relations is the main subject of weak measurement~\cite{aharonov2}. We are thus motivated for an examination of the interplay of entanglement and measurement-disturbance in Hardy's model.

In quantum mechanics, entanglement and uncertainty principle are two logically independent notions in the sense that entanglement can in principle be quantified without 
spoiling uncertainty principle; for example, using Heisenberg's uncertainty relation in the manner of Kennard~\cite{kennard} one can derive a general necessary
condition for separability of two-party systems with one continuous freedom in
each party~\cite{fujikawa1}. 
This condition, which is valid for general mixed states, leads to the well-known necessary and sufficient separability condition when applied to two-party Gaussian systems in quantum optics~\cite{duan}. In contrast,
Bell-CHSH inequalities are based on hidden-variables models (local realism), which do not satisfy all the properties of quantum mechanics, and thus it is not obvious if uncertainty principle is preserved in such models. Hardy's model also
depends on hidden-variables models in an essential manner, and thus his model could 
spoil uncertainty principle. In fact, we are going to show that Hardy's model induces the breaking of uncertainty principle in the sense of measurement-disturbance relations in the conditional measurement of non-commuting operators. It is, however, important that his analysis of the inconsistency of local realism with entanglement is simplified if this breaking of measurement-disturbance relations is taken into account, and a much simplified experimental test of the conflict of local realism with quantum mechanics is illustrated using Hardy's thought experiment. A general difficulty of hidden-variables models to describe the conditional measurement of non-commuting operators is also illustrated using the simplest hidden-variables model in $d=2$ proposed by Bell~\cite{bell2}.  
 
In our main analysis, we use the hidden-variables model in $d=4$ (local realism) that belongs to the same class of models as in the original paper of Bell~\cite{bell} which is non-contextual and local (i.e., a non-contextual model applied to two far-apart parties)\cite{mermin}, although  due care needs to be exercised in the treatment of non-contextual hidden-variables models in the Hilbert space with dimension $d=4$~\cite{kochen, beltrametti}.

\section{Thought experiment}
We start with a review of Hardy's thought experiment~\cite{hardy1}. The state (eq.(9) in~\cite{hardy1}),  
\begin{eqnarray}
|\psi\rangle=\frac{1}{2}\left(-|\gamma\rangle +i|u^{+}\rangle|v^{-}\rangle +i|v^{+}\rangle|u^{-}\rangle +|v^{+}\rangle|v^{-}\rangle\right),
\end{eqnarray}
defines an entangled state generated by a quantum mechanical elimination of $|u^{+}\rangle|u^{-}\rangle$ from a separable state $(1/\sqrt{2})(i|u^{+}\rangle+|v^{+}\rangle)(1/\sqrt{2})(i|u^{-}\rangle+|v^{-}\rangle)$. The physical meanings of various states here are that  $(1/\sqrt{2})(i|u^{+}\rangle+|v^{+}\rangle)$ stands for the positron state after going through a beam splitter into two paths specified by $u^{+}$ and $v^{+}$ and, similarly, the state $(1/\sqrt{2})(i|u^{-}\rangle+|v^{-}\rangle)$ stands for the electron state after passing through another beam splitter. The missing state $|u^{+}\rangle|u^{-}\rangle$ is assumed to have pair annihilated into a gamma ray $|\gamma\rangle$. 

The notion of {\em locality} is implemented by assuming that the electron and positron can interact only in  the state $|u^{+}\rangle|u^{-}\rangle$  to annihilate into a gamma ray to generate an entangled electron-positron state, but except for this interaction the electron and positron never come close to each other and their later time development is  completely independent.
  
Further applications of beam splitters in~\cite{hardy1} introduce  unitary transformations in each party separately and thus provide various {\em detector states} for each particle separately; $C^{-}(\infty)=|u^{-}\rangle\langle u^{-}|$ and $D^{-}(\infty)=|v^{-}\rangle\langle v^{-}|$  are transformed to
\begin{eqnarray}
&&C^{-}(0)=|(u^{-}-iv^{-})/\sqrt{2}\rangle\langle (u^{-}+iv^{-})/\sqrt{2}|, \nonumber\\
&&D^{-}(0)=|(v^{-}-iu^{-})/\sqrt{2}\rangle\langle (v^{-}+iu^{-})/\sqrt{2}|
\end{eqnarray}
for the electron, and $C^{+}(\infty)=|u^{+}\rangle\langle u^{+}|$ and $D^{+}(\infty)=|v^{+}\rangle\langle v^{+}|$  are transformed to
\begin{eqnarray}
&&C^{+}(0)=|(u^{+}-iv^{+})/\sqrt{2}\rangle\langle (u^{+}+iv^{+})/\sqrt{2}|, \nonumber\\
&&D^{+}(0)=|(v^{+}-iu^{+})/\sqrt{2}\rangle\langle (v^{+}+iu^{+})/\sqrt{2}|
\end{eqnarray}
for the positron, respectively.  But the physical state is always given by the initial state $|\psi\rangle$ in (1). The locality requirement means that $C^{-}(0)$ and $D^{-}(0)$ in (2) are completely independent of  $C^{+}(0)$ and $D^{+}(0)$ in (3).  The measurements are performed by applying various detector states to the entangled initial state  $|\psi\rangle$.  The initial state $|\psi\rangle$ in (1) is rewritten in terms of various different sets of states in~\cite{hardy1}, but the physical content is the same as the state $|\psi\rangle$ in (1).  Note that the notion of reduction, namely, the change of the state by measurement is not incorporated in a theory based on local realism.

Starting with (1), we find in the  notation of Hardy with $\lambda$ standing for hidden-variables~\cite{hardy1},
\begin{eqnarray}
&&C^{+}(\infty,\lambda)C^{-}(\infty,\lambda)=0,\nonumber\\
&&C^{+}(\infty,\lambda)=1 \Rightarrow D^{-}(\infty,\lambda)=1,\nonumber\\
&&C^{-}(\infty,\lambda)=1 \Rightarrow D^{+}(\infty,\lambda)=1,\nonumber\\
&&D^{+}(\infty,\lambda)D^{-}(\infty,\lambda)>0.
\end{eqnarray}
 The bold rightarrow indicates "inevitably implies" or "their existence is inferred on the basis of predictions
of probability equal to 1". The general notational conventions are: $C^{\pm}(\infty,\lambda)$ and $C^{\pm}(0,\lambda)$ are the measured results (in hidden-variables models) of projection operators $C^{\pm}(\infty)=|u^{\pm}\rangle\langle u^{\pm}|$ and $C^{\pm}(0)$ in (2) and (3) for the state $|\psi\rangle$ in (1), respectively.
 Similarly, $D^{\pm}(\infty,\lambda)$ and $D^{\pm}(0,\lambda)$ specify the measured results of projection operators $D^{\pm}(\infty)=|v^{\pm}\rangle\langle v^{\pm}|$ and $D^{\pm}(0)$ in (2) and (3) for the state $|\psi\rangle$ in (1), respectively.

The first relation $C^{+}(\infty,\lambda)C^{-}(\infty,\lambda)=0$ in (4), for example, shows that we have no state $|u^{+}\rangle|u^{-}\rangle$ in (1). The second relation $C^{+}(\infty,\lambda)=1 \Rightarrow D^{-}(\infty,\lambda)=1$ in (4) is understood by $(|u^{+}\rangle)^{\dagger}|\psi\rangle=(i/2)|v^{-}\rangle$, where the left-hand side shows the measurement of $|\psi\rangle$ by $C^{+}(\infty)$ and the right-hand side shows the state to be measured by $D^{-}(\infty)$.  It is confirmed that the relations in (4) do not lead to any paradox.
 
Following the analysis in~\cite{hardy1}, we further find
\begin{eqnarray}
&&D^{-}(0,\lambda)=1 \Rightarrow C^{+}(\infty,\lambda)=1 \Rightarrow D^{-}(\infty,\lambda)=1, 
\end{eqnarray}
where we used  the relation (4) in the second step, and similarly
\begin{eqnarray}
&&D^{+}(0,\lambda)=1 \Rightarrow C^{-}(\infty,\lambda)=1 \Rightarrow D^{+}(\infty,\lambda)=1. 
\end{eqnarray}
The relation (5) is confirmed by noting $[(|v^{-}\rangle-i|u^{-}\rangle)/\sqrt{2}]^{\dagger}|\psi\rangle=(i/2\sqrt{2})|u^{+}\rangle$ where the left-hand side shows the measurement of the state $|\psi\rangle$ by $D^{-}(0)$ and the right-hand side shows the state to be measured by $C^{+}(\infty)$, and similarly the relation (6) is confirmed. It is known that $C^{+}(\infty,\lambda)C^{-}(\infty,\lambda)=0$ in (4) and the relations (5) and (6) combined with $D^{-}(0,\lambda)D^{+}(0,\lambda)\neq 0$ give rise to a paradox~\cite{hardy1}.

The relation (5), $D^{-}(0,\lambda)=1 \Rightarrow D^{-}(\infty,\lambda)=1$, 
or the corresponding hidden-variables representation
\begin{eqnarray}
\frac{\int D^{-}(\infty,\lambda)D^{-}(0,\lambda)d\mu(\lambda)}{\int D^{-}(0,\lambda)d\mu(\lambda)}=1,
\end{eqnarray}
when translated into the quantum mechanical language, implies a statement
\begin{eqnarray}
\frac{\langle\psi| D^{-}(0)D^{-}(\infty)D^{-}(0)|\psi\rangle}{
\langle\psi| D^{-}(0)|\psi\rangle}=1,
\end{eqnarray}
which is a contradiction, since $D^{-}(0)D^{-}(\infty)D^{-}(0)=cD^{-}(0)$
and thus 
\begin{eqnarray}
\frac{\langle\psi| D^{-}(0)D^{-}(\infty)D^{-}(0)|\psi\rangle}{
\langle\psi| D^{-}(0)|\psi\rangle}=c,
\end{eqnarray}
with
\begin{eqnarray}
c={\rm Tr}D^{-}(0)D^{-}(\infty)=\frac{1}{2}<1,
\end{eqnarray}
in the present case. The prediction (8) is another  logical inconsistency (paradox) between quantum mechanics and local realism. 

The prediction (5) of local realism is also written as 
\begin{eqnarray}
D^{-}(\infty,\lambda)=0 \ \Rightarrow C^{+}(\infty,\lambda)=0 \ \Rightarrow D^{-}(0,\lambda)=0.
\end{eqnarray}
 Namely,
the null result of $D^{-}(\infty)=|v^{-}\rangle\langle v^{-}|$ before the beam splitter which determines the state to be measured, $D^{-}(0)$ or $C^{-}(0)$ in (2), inevitably implies the null result $D^{-}(0,\lambda)=0$ for the state $|\psi\rangle$ in (1). Obviously, this is wrong in quantum mechanics due to the reduction of states by measurement, as is seen in the state $|\psi\rangle$ in (1); the null result of $D^{-}(\infty)=|v^{-}\rangle\langle v^{-}|$ implies the projection of the state
\begin{eqnarray}
|\psi\rangle \rightarrow (1-D^{-}(\infty))|\psi\rangle=\frac{1}{2}\left(-|\gamma\rangle +i|v^{+}\rangle|u^{-}\rangle\right)\nonumber 
\end{eqnarray}
which obviously has non-vanishing overlap with $D^{-}(0)$ in (2).
This may be experimentally tested using the technique of weak measurement~\cite{aharonov1}. Theoretically, an equivalent statement of (11) is that $\bar{D}^{-}(\infty,\lambda)=1-D^{-}(\infty,\lambda)=1 \Rightarrow \bar{D}^{-}(0,\lambda)=1-D^{-}(0,\lambda)=1$, namely,
\begin{eqnarray}
\frac{\langle\psi| \bar{D}^{-}(\infty)\bar{D}^{-}(0)\bar{D}^{-}(\infty)|\psi\rangle}{
\langle\psi| \bar{D}^{-}(\infty)|\psi\rangle}=1,
\end{eqnarray}
while quantum mechanically the right-hand side of (12) should be ${\rm Tr}\bar{D}^{-}(0)\bar{D}^{-}(\infty)=1/2$, and this provides a way alternative to (8) to realize the inconsistency with quantum mechanics. It is important to recognize that both of these tests, (8) and (12), are based on the consideration of only two-steps of measurements instead of four-steps in the original analysis of Hardy.

The relation (9) is regarded as a statement of measurement-disturbance relation in uncertainty principle; any state is projected to the eigenstate of the projector such as $D^{-}(0)$ if one measures it, and the probability of $D^{-}(\infty)$ with $[D^{-}(\infty), D^{-}(0)]\neq 0$ for the new state inevitably deviates from unity.  The quantity $c$ in (10) is a measure of the disturbance in the eigenstate of $D^{-}(\infty)$, which satisfies $\langle\psi| D^{-}(\infty)|\psi\rangle=1$, induced by the measurement of projector  $D^{-}(0)$. To be explicit, using the state 
\begin{eqnarray}
\rho_{f}=D^{-}(0)|\psi\rangle\langle\psi|D^{-}(0)/\langle\psi|D^{-}(0)|\psi\rangle
\end{eqnarray}
after the measurement of $D^{-}(0)=|(v^{-}-iu^{-})/\sqrt{2}\rangle\langle (v^{-}+iu^{-})/\sqrt{2}|$, the standard deviation of $D^{-}(\infty)=|v^{-}\rangle\langle v^{-}|$ is given by 
\begin{eqnarray}
\sigma_{f}(D^{-}(\infty))&=&[{\rm Tr}\rho_{f}D^{-}(\infty)^{2}- \left({\rm Tr}\rho_{f}D^{-}(\infty)\right)^{2}]^{1/2}\nonumber\\
&=&\sqrt{c(1-c)}, 
\end{eqnarray}
while $\sigma_{i}(D^{-}(\infty))=0$
for the initial state with $ D^{-}(\infty)|\psi\rangle=|\psi\rangle$. We thus call the relation (9) as a manifestation of the measurement-disturbance relation in the present study. We also note that the quantity, $-\log c$, characterizes the lower bound of entropic uncertainty relation~\cite{uffink} which is state-independent; note that the original relation of Heisenberg,
$\delta q \delta p \sim \hbar$, is state-independent. An interesting analysis of the relation between entropic uncertainty relations and non-locality is found  in~\cite{oppenheim}.

Starting with a system without any paradox (4), one generates a paradox related to entanglement~\cite{hardy1} by local operations. From a point of view of local parties, those local operations lead to the violation of  measurement-disturbance relations in the sense of  (8) or (12) in hidden-variables models (local realism).

\section{Hardy's model}

We next recapitulate the  concrete 
model of Hardy~\cite{hardy2} and show in a more explicit manner that the conflict with uncertainty relations in the sense of measurement-disturbance relations is involved. The model consists of the physical projection operators
\begin{eqnarray}
U_{i}=|u_{i}\rangle\langle u_{i}|,\ \
D_{i}=|d_{i}\rangle\langle d_{i}|,
\end{eqnarray}
with $i=1,2$, and 
\begin{eqnarray}
|u_{i}\rangle&=&\frac{1}{\sqrt{\alpha+\beta}}[\beta^{1/2}|+\rangle_{i}+\alpha^{1/2}|-\rangle_{i}],\nonumber\\
|d_{i}\rangle&=&\frac{1}{\sqrt{\alpha^{3}+\beta^{3}}}[\beta^{3/2}|+\rangle_{i}-\alpha^{3/2}|-\rangle_{i}]
\end{eqnarray}
for the entangled state of two far-apart qubits, 
\begin{eqnarray}
|\psi\rangle=\alpha|+\rangle_{1}|+\rangle_{2}-\beta|-\rangle_{1}|-\rangle_{2}
\end{eqnarray}
with $\alpha^{2}+\beta^{2}=1$. The  concrete physical meaning of these qubits is not important. 
 Note that the state $|\psi\rangle$ here has no direct connection with the state in (1), although we use the same generic notation  for a state vector.
We work with real and non-negative $\alpha$ and $\beta$, for simplicity, but more general cases can be treated similarly. For $\alpha\neq \beta$, we have $[D_{1},U_{1}]\neq 0$ and $[D_{2},U_{2}]\neq 0$.

We then obtain the relations (note that $D_{1}U_{2}D_{1}=D_{1}U_{2}$, for example)
\begin{eqnarray}
\frac{\langle\psi|D_{1}U_{2}D_{1}|\psi\rangle}{\langle\psi|D_{1}|\psi\rangle}&=&1,\\
\frac{\langle\psi|D_{2}U_{1}D_{2}|\psi\rangle}{\langle\psi|D_{2}|\psi\rangle}&=&1,\\
\frac{\langle\psi|D_{1}D_{2}D_{1}|\psi\rangle}{\langle\psi|D_{1}|\psi\rangle}&=&\frac{(\beta-\alpha)^{2}}{(\beta-\alpha)^{2}+\beta\alpha},
\\
\langle\psi|U_{1}U_{2}|\psi\rangle=0,
\end{eqnarray}
and $\langle\psi|D_{1}|\psi\rangle=\langle\psi|D_{2}|\psi\rangle
=\alpha^{2}\beta^{2}/(1-\alpha\beta)$
with $0< \alpha\beta\leq 1/2$.

Hardy then argues that~\cite{hardy2}:\\
i) The measured value of $\langle D_{1}D_{2}D_{1}\rangle=\langle D_{1}D_{2}\rangle\neq 0$ in (20) for $0< \alpha\beta< 1/2$ (i.e., $\alpha \neq \beta$) implies 
$D_{1}(\psi, \lambda)=D_{2}(\psi, \lambda)=1$ for {\em some} $\lambda \in \Lambda$, which is based on the  assumption that the non-contextual and local hidden-variables model (local realism)
\begin{eqnarray}
\langle\psi| D_{1}D_{2}|\psi\rangle=\int_{\Lambda} d\mu(\lambda) D_{1}(\psi, \lambda)D_{2}(\psi, \lambda),
\end{eqnarray}
is valid for this combination. As for the notational convention, $D_{1}(\psi, \lambda)$, for example, stands for the eigenvalues of the projection operator $D_{1}$, namely, $1$ or $0$, depending on the hidden-variables $\lambda$. We write the state $\psi$ dependence explicitly such as in $D_{1}(\psi, \lambda)$ in conformity with the explicit $d=2$ model to be discussed later; this  $\psi$ dependence is implicit in the notation such as $D^{-}(0,\lambda)$ of Section 2 which follows the notation of Hardy.  This difference does not modify the conclusion.\\
ii) The assumption  of the validity of the conditional probability
\begin{eqnarray}
\frac{\langle\psi|D_{2}U_{1}D_{2}|\psi\rangle}{\langle\psi|D_{2}|\psi\rangle}&=&\frac{\int_{\Lambda} d\mu(\lambda) U_{1}(\psi, \lambda)D_{2}(\psi, \lambda)
}{\int_{\Lambda} d\mu(\lambda) D_{2}(\psi, \lambda)}\nonumber\\
&=&1,
\end{eqnarray}
leads to  $D_{2}(\psi, \lambda)=1 \Rightarrow U_{1}(\psi, \lambda)=1$.\\
iii) Similarly, the assumption  of the validity of  the conditional probability
\begin{eqnarray}
\frac{\langle\psi|D_{1}U_{2}D_{1}|\psi\rangle}{\langle\psi|D_{1}|\psi\rangle}&=&\frac{\int_{\Lambda} d\mu(\lambda) U_{2}(\psi, \lambda)D_{1}(\psi, \lambda)
}{\int_{\Lambda} d\mu(\lambda) D_{1}(\psi, \lambda)}\nonumber\\&=&1,
\end{eqnarray}
leads to  $D_{1}(\psi, \lambda)=1 \Rightarrow U_{2}(\psi, \lambda)=1$.\\
iv) The assumption
\begin{eqnarray}
\langle\psi| U_{1}U_{2}|\psi\rangle=\int_{\Lambda} d\mu(\lambda) U_{1}(\psi, \lambda)U_{2}(\psi, \lambda),
\end{eqnarray}
then implies that $\langle\psi| U_{1}U_{2}|\psi\rangle\neq 0$, but this contradicts the prediction of quantum mechanics $\langle\psi| U_{1}U_{2}|\psi\rangle= 0$ in (21).  This contradiction is commonly referred to as "Hardy's paradox".
This analysis demonstrates,  without referring to CHSH inequalities~\cite{bell, chsh}, that local realism (non-contextual and local hidden-variables model) cannot explain  entanglement~\cite{hardy2}.

To show the conflict with measurement-disturbance relations in the present model, we consider only the last three processes (ii), (iii) and (iv). The relation $\langle\psi| U_{1}U_{2}|\psi\rangle= 0$ in (21) then implies 
\begin{eqnarray}
U_{1}(\psi, \lambda)U_{2}(\psi, \lambda)=0
\end{eqnarray}
in the context of (25).
One can easily confirm that the three processes (ii), (iii) and (iv) are consistent in the sense of Hardy with emphasis on entanglement without the condition coming from (i). 
But one recognizes the following two new conditions from (ii), (iii) and (iv),
\begin{eqnarray}
&&D_{1}(\psi, \lambda)=1 \Rightarrow U_{1}(\psi, \lambda)=0, \nonumber\\
&&D_{2}(\psi, \lambda)=1 \Rightarrow U_{2}(\psi, \lambda)=0,
\end{eqnarray}
since the relation (26) shows that $U_{2}(\psi, \lambda)=1$ implies $U_{1}(\psi, \lambda)=0$, and similarly, $U_{1}(\psi, \lambda)=1$ implies $U_{2}(\psi, \lambda)=0$.
The first relation of (27) implies  a quantum mechanical relation
$\langle \psi|D_{1}U_{1}D_{1}|\psi\rangle/\langle\psi| D_{1}|\psi\rangle=0$ or equivalently,
\begin{eqnarray}
\frac{\langle\psi| D_{1}\bar{U}_{1}D_{1}|\psi\rangle}{\langle\psi| D_{1}|\psi\rangle}=1,
\end{eqnarray}
if one defines $ \bar{U}_{1}\equiv 1-U_{1}$. This prediction contradicts the 
quantum mechanical prediction
\begin{eqnarray}
\frac{\langle \psi|D_{1}\bar{U}_{1}D_{1}|\psi\rangle}{\langle\psi| D_{1}|\psi\rangle}=\bar{c},
\end{eqnarray}
where
\begin{eqnarray}
\bar{c}={\rm Tr}\bar{U}_{1}D_{1}=1-(\beta-\alpha)^{2}/[(\beta-\alpha)^{2}+\beta\alpha]
\end{eqnarray}
with $0<\bar{c}<1$ for $\alpha\neq \beta$ which is required in Hardy's analysis~\cite{hardy2}. We thus recognize the conflict of local realism with  measurement-disturbance relations  without directly referring to the conflict with entanglement, although the correlations between two parties such as in (ii), (iii) and (iv) are the consequences of entanglement. 

It should be emphasized that we need to consider only two processes, (iii) and (iv), to recognize the conflict of local realism with quantum mechanics (paradox) instead of four required in the original scheme of Hardy. This is related to the simplified experimental test of the prediction of local realism in (11) in the previous section. The basic mechanism of this simplified test is clarified in the next section.

\section{Conditional measurement }
The examples of the violation of uncertainty relations we mentioned are related to the analysis of measurement-disturbance relations in conditional measurement.
We thus start with a discussion of conditional measurement  
in the well-defined $d=2$ hidden-variables model proposed by Bell~\cite{bell2, beltrametti}, since the entire subject of Bell-CHSH inequalities started when Bell
recognized this hidden-variables model in $d=2$. 
To analyze the conditional measurement, we introduce some new notations.     
The probability measure associated with the projection operator $A$ and  the state $|\psi\rangle$ is defined by 
\begin{eqnarray}
\mu[a_{\psi}]\equiv \int_{\Lambda}A_{\psi}(\lambda)d\mu(\lambda) =
\langle\psi|A|\psi\rangle,
\end{eqnarray}
with the auxiliary quantity  $a_{\psi}$ defined by 
$a_{\psi}=\{\lambda\in \Lambda : A_{\psi}(\lambda)=1\}$.

The quantity $A_{\psi}(\lambda)$ which depends on the hidden-variables $\lambda$ is also written as  $A(\psi,\lambda)$ in the present paper.
The basic assumption in hidden variables models is that one can find 
$A_{\psi}(\lambda)$, which assumes the eigenvalues of the projector $A$, namely, $1$ or $0$,  and a measure $\mu[a_{\psi}]$ to reproduce the quantum mechanical $\langle\psi|A|\psi\rangle$ for any $A$ and $|\psi\rangle$. 
Bell's non-contextual hidden-variables model in $d=2$ is defined by~\cite{bell2, beltrametti}  
\begin{eqnarray}
A_{\psi}(\lambda)=\frac{1}{2}[1+{\rm sign}(\lambda+\frac{1}{2}|{\vec{s}}\cdot {\vec{m}}|){\rm sign}({\vec{s}}\cdot {\vec{m}})],
\end{eqnarray}
for the projection operator 
$A \equiv P_{\vec{m}}=\frac{1}{2}(1+\vec{m}\cdot\vec{\sigma})$ 
with a unit vector $\vec{m}$ and the Pauli matrix $\vec{\sigma}$, and the initial pure state given by 
$\rho=|\psi\rangle\langle\psi|=\frac{1}{2}(1+\vec{s}\cdot\vec{\sigma})$ with a unit vector $\vec{s}$. This concrete construction of $A_{\psi}(\lambda)$, which assumes the eigenvalues of the projection operator $A$, $1$ or $0$, explicitly depends on the initial 
state $\rho$ and the hidden-variable $\lambda$; following this explicit example, we write the state dependence explicitly as in $A_{\psi}(\lambda)$ in the present paper.
It is then confirmed that 
$\int_{-1/2}^{1/2}d\lambda A_{\psi}(\lambda)=\langle\psi|A|\psi\rangle$
with a non-contextual weight factor $d\mu(\lambda)=d\lambda$ which is independent of $A$; in the present case, the weight happens to be uniform. A general hermitian operator $O$ is treated by
performing the spectral decomposition $O=\sum_{k}\mu_{k}P_{k}$ with orthogonal projectors $P_{k}$. For example, $\vec{a}\cdot\vec{\sigma}=\sum_{k}\mu_{k}P_{k}$ which justifies the use of dichotomic variables ($\pm 1$) for $\vec{a}\cdot\vec{\sigma}$.

The quantum mechanical conditional probability of the measurement of a projector $B$ after the measurement of a projector $A$ is defined by~\cite{ davies, fujikawa2}
\begin{eqnarray}
P(B|A)=\frac{\langle\psi|ABA|\psi\rangle}{\langle\psi|A|\psi\rangle}
\end{eqnarray}
for $\langle\psi|A|\psi\rangle\neq 0$. One may evaluate $P(B|A)$ by
starting with the separate hidden-variables constructions of $A$ and $B$.
For non-commuting $A$ and $B$, however, $(ABA)_{\psi}(\lambda)\neq A_{\psi}(\lambda)B_{\psi}(\lambda)A_{\psi}(\lambda)=A_{\psi}(\lambda)B_{\psi}(\lambda)$ in general, where the left-hand side stands for a single positive operator  while the right-hand side stands for a product of projectors.
 The classical conditional probability rule (as in non-contextual hidden-variables models) for non-commuting $A$ and $B$ is given by Bayes rule, namely, by the left-hand side of 
\begin{eqnarray}
\frac{\mu[b_{\psi}\cap a_{\psi}]}{\mu[a_{\psi}]}=\frac{\langle\psi|ABA|\psi\rangle}{\langle\psi|A|\psi\rangle}.
\end{eqnarray}
The classical conditional probability rule may give rise to the quantum probability on the right-hand side for some specific case by choosing a clever representation such as $A_{\psi}(\lambda)$  and the weight $d\mu(\lambda)$, but not in general; one can confirm that the explicit example in Bell's model in (32) does not satisfy this condition, namely, 
\begin{eqnarray}
P(B|A)\neq \int d\mu(\lambda)A_{\psi}(\lambda)B_{\psi}(\lambda)/\int d\mu(\lambda)A_{\psi}(\lambda).
\end{eqnarray}
Also, it is shown that 
the relation (34), if imposed for {\em all} $|\psi\rangle$, implies commuting $A$ and $B$~\cite{malley}; this conclusion is valid for any dimension $d$. 

It is thus not surprising that hidden-variables models have a difficulty in describing conditional probability in general as we have recognized by analyzing Hardy's model. A salient feature of Hardy's model is that we can make a definite statement independent of detailed specifications of hidden-variables models. 

The basic reason for the failure of classical conditional probability  (34) is traced to the absence of reduction, namely, $b_{\psi}$ and $a_{\psi}$  are determined by  the {\em same} state $|\psi\rangle$. 
It is instructive to re-examine the conditional probability $P(U_{2}|D_{1})$ (first measure $D_{1}$ and then measure $U_{2}$) in Hardy's model from the point of view of conditional measurement. We start with  (24)
\begin{eqnarray} 
&&\frac{\langle\psi|(D_{1} \otimes 1)(1\otimes U_{2})(D_{1}\otimes 1)|\psi\rangle}
{\langle\psi|(D_{1}\otimes 1)|\psi\rangle}\nonumber\\
&&=\frac{\int d\mu(\lambda) D_{1}(\psi,\lambda)U_{2}(\psi,\lambda)}
{\int d\mu(\lambda)D_{1}(\psi,\lambda)}=\frac{ \mu[u_{2\psi}\cap d_{1\psi}]}{\mu[d_{1\psi}]}.
\end{eqnarray}
This classical probability rule, although the necessary condition $[(D_{1}\otimes 1),(1\otimes U_{2})]=0$ for the relation (34) to be valid for any $|\psi\rangle$ is satisfied,  cannot describe entanglement since the measurement 
of the projector $D_{1}$ disturbs the state $|\psi\rangle$ and thus the subsequent measurement of 
$U_{2}$ is determined by the modified state if the state $|\psi\rangle$ is entangled; the above probability rule contains $u_{2\psi}$ which is determined by the {\em initial state} $|\psi\rangle$ and thus cannot incorporate the effects of entanglement and reduction in general. 

This analysis (36) is associated with Hardy's paradox in the following manner. One may write (21) as $P(U_{1}|U_{2})=0$, namely,
\begin{eqnarray}
\langle\psi|(1\otimes U_{2})( U_{1}\otimes 1)(1\otimes U_{2})|\psi\rangle/
\langle\psi|(1\otimes U_{2})|\psi\rangle=0,
\end{eqnarray}
which implies $U_{2}(\psi,\lambda)=1 \Rightarrow U_{1}(\psi,\lambda)=0$ with the same state $|\psi\rangle$ in hidden-variables models. Similarly, the relation  $P(U_{2}|D_{1})=1$ in (24) (and in (36)) implies $D_{1}(\psi,\lambda)=1 \Rightarrow U_{2}(\psi,\lambda)=1$ in hidden-variables models. These two relations put together implies $D_{1}(\psi,\lambda)=1 \Rightarrow U_{1}(\psi,\lambda)=0$ with the same state $|\psi\rangle$, but this assigns definite eigenvalues to two non-commuting operators  $D_{1}$ and $U_{1}$ for the same state $|\psi\rangle$ simultaneously. This is a contradiction with quantum mechanics, which 
we have characterized by the expression (28) that violates the measurement-disturbance relation  in (29). 

The essence of Hardy's model is that (24) and (25), to be precise (18) and (21), are two conditional measurements $P(U_{2}|D_{1})=1$ and $P(U_{1}|U_{2})=0$, which give rise to definite eigenvalues to two non-commuting $D_{1}$ and $U_{1}$ for $\alpha\neq \beta$  simultaneously in hidden-variables models (local realism); we have $[D_{1},U_{1}]\neq 0$ except maximally entangled states, which explains why Hardy's paradox does not appear for maximally entangled states.
 
\section{Conclusion} 

It has been shown that the formulation of Hardy's paradox inevitably
contains the violation  of  measurement-disturbance relations.
Both of Hardy's paradox and the failure of  measurement-disturbance relations we discussed arise from the absence of the notion of state reduction in non-contextual and local hidden-variables models (local realism). Hardy's paradox motivated many experimental works and it is currently actively discussed in the context of weak measurement, and the analysis of general aspects of measurement-disturbance is the main subject of weak measurement. 
It was emphasized that the analysis of the inconsistency of local realism with entanglement by Hardy is simplified if the breaking of  measurement-disturbance relations is taken into account, and a much simplified test of the conflict of local realism with measurement-disturbance relations, which may be performed by weak measurement, is illustrated in the framework of Hardy's thought experiment. The basic mechanism of this simplified test was further clarified by a detailed analysis of conditional measurement by showing that the simplest hidden-variables model of Bell in $d=2$~\cite{bell2}, which has been believed to reproduce full quantum mechanics, 
already fails to describe conditional measurement.

In connection with uncertainty principle itself, 
it has been recognized recently that the characterization of Heisenberg's error-disturbance relation by inequalities is more involved than  hitherto assumed~\cite{ozawa, werner, ueda}, although Heisenberg's idea of the measurement-disturbance of two non-commuting operators, as we utilized in the present paper, is believed to be valid. In view of these current developments in the fundamental aspects of quantum mechanics, our analysis of  Hardy's paradox with emphasis on the conditional measurement and measurement-disturbance should be relevant.
 Hardy's paradox is resolved for the maximally entangled state, for which specific operators in his model become commuting and thus the difficulty associated with measurement-disturbance relations disappears. 
 
It should be useful to keep the intimate interplay of entanglement and measurement-disturbance in mind when one appreciates  past experiments~\cite{togerson} - \cite{vallone} and contemplates further experimental tests of Hardy's paradox, in particular, in connection with weak measurement.

\section*{Acknowledgments}  
One of us (K.F.) thanks the hospitality at the Center for Quantum Technologies. This work is partially supported by the National Research Foundation and Ministry of Education, Singapore (Grant No. WBS: R-710-000-008-271) and JSPS KAKENHI (Grant No. 25400415).


\begin{thebibliography}{99}
\bibitem{hardy1}
L. Hardy, Phys. Rev. Lett. {\bf 68}, 2981 (1992).
\bibitem{hardy2}
L. Hardy, Phys. Rev. Lett. {\bf 71}, 1665 (1993).
\bibitem{bell}
J. S. Bell, Physics {\bf 1}, 195 (1965).
\bibitem{chsh}
J. F. Clauser, M. A. Horne, A. Shimony and R. A. Holt, Phys. Rev. Lett. 
{\bf 23},  880 (1969). 
\bibitem{togerson}
J. R. Togerson, D. Branning, C. H. Monken and L. Mandel, Phys. Lett. A{\bf 204}, 323 (1995).
\bibitem{giuseppe}
G. Di Giuseppe, F. De Martini, and D. Boschi, Phys. Rev. A{\bf 56}, 176 (1997).
\bibitem{white}
A. G. White, D. F. V. James, P. H. Eberhard and P. G. Kwiat, Phys. Rev. Lett., {\bf 83}, 3103 (1999).
\bibitem{Irvine} 
W. T. M. Irvine, J. F. Hodelin, C. Simon, and D. Bouwmeester, Phys. Rev. Lett. {\bf 95}, 030401 (2005).
\bibitem{carlson}
J. A. Carlson, M. D. Olmstead and M. Beck, Am. J. Phys., {\bf 74}, 180 (2006).
\bibitem{lundeen}
J. S. Lundeen and A. M. Steinberg, Phys. Rev. Lett. {\bf 102}, 020404 (2009).
\bibitem{imoto}
K. Yokota, T. Yamamoto, M. Koashi, and N. Imoto, New J. Phys. {\bf 11}, 033011  (2009).
\bibitem{fedrizzi}
A. Fedrizzi, M. P. Almeida, M. A. Broome, A. G. White, and M. Barbieri, Phys. Rev. Lett. {\bf 106}, 200402 (2011).
\bibitem{vallone}
G. Vallone, I. Gianani, E. B. Inostroza, C. Saavedra, G. Lima, A. Cabello, and P. Mataloni, Phys. Rev. A{\bf 83}, 042105 (2011).
\bibitem{aharonov1}
Y. Aharonov, A. Botero, S. Popescu, B. Reznik,
J. Tollaksen, Phys. Lett. A{\bf 301}, 130 (2002).
\bibitem{aharonov2}
Y. Aharonov and L. Vaidman, Phys. Rev. A{\bf 41}, 11 (1990).
\bibitem{kennard}
E.H. Kennard, Z. Phys. {\bf 44}, 326 (1927).
\bibitem{fujikawa1}
K. Fujikawa, Phys. Rev. A{\bf 80}, 012315 (2009).
\bibitem{duan}
L.M. Duan, G. Giedke, J.I. Cirac and P. Zoller, Phys. Rev. Lett. {\bf 84}, 2722 (2000). \\
R. Simon, Phys. Rev. Lett. {\bf 84}, 2726 (2000).
\bibitem{bell2}
J. S. Bell, Rev. Mod. Phys. {\bf 38}, 447 (1966).
\bibitem{mermin}
See N. D. Mermin, Rev. Mod. Phys. {\bf 65}, 803 (1993), for a  classification of hidden-variables models.
\bibitem{kochen}
S. Kochen and E. P. Specker, J. Math. Mech. {\bf 17}, 59 (1967).
\bibitem{beltrametti}
E. G. Beltrametti and G. Gassinelli, {\em The Logic of Quantum 
Mechanics}, (Addison-Wesley Pub., 1981).
\bibitem{uffink}
H. Maassen and J. B. M. Uffink, Phys. Rev. Lett. {\bf 60}, 1103 (1988).
\bibitem{oppenheim}
J. Oppenheim and S. Wehner, Science {\bf 330}, 1072 (2010).
\bibitem{davies}
E. B. Davies and J. T. Lewis, Comm. Math. Phys. {\bf 17}, 239 (1970).  
\bibitem{fujikawa2}
K. Fujikawa, Phys. Rev. A{\bf 85}, 012114 (2012).
\bibitem{malley}
J. D. Malley and A. Fine, Phys. Lett. A{\bf 347}, 51 (2005).
\bibitem{ozawa}
J. Erhart, S. Sponar, G. Sulyok, G. Badurek, M. Ozawa, and Y. Hasegawa, 
Nat.\ Phys.\ \textbf{8}, 185, (2012).
\bibitem{werner}
P. Busch, P. Lahti, and R.F. Werner, Phys. Rev. Lett.
111, 160405 (2013).
\bibitem{ueda}
Y. Watanabe, T. Sagawa, and M. Ueda, Phys. Rev. A{\bf 84}, 042121 (2011).\\
Y. Watanabe and M. Ueda, arXiv:1106.2526 [quant-ph] (2011).\\
K. Fujikawa, Phys. Rev. A{\bf 88}, 012126 (2013), and references therein.
\end{thebibliography}
\end{document}